\documentclass[,final]{aipproc}
\layoutstyle{6x9}

  \makeatletter
  \def\mathcomposite{%
     \@ifstar
        {\def\@mathcomposite@option{%
            \baselineskip\z@skip\lineskiplimit-\maxdimen}%
         \@mathcomposite}%
        {\let\@mathcomposite@option\offinterlineskip
         \@mathcomposite}}
  \def\@mathcomposite{%
     \@ifnextchar[\@@mathcomposite{\@@mathcomposite[0]}}
  \def\@@mathcomposite[#1]#2#3#4{%
     #2{\mathchoice
        {\@mathcomposite@{#1}{#3}{#4}\displaystyle{1}}%
        {\@mathcomposite@{#1}{#3}{#4}\textstyle{1}}%
        {\@mathcomposite@{#1}{#3}{#4}%
         \scriptstyle\defaultscriptratio}%
        {\@mathcomposite@{#1}{#3}{#4}%
         \scriptscriptstyle\defaultscriptscriptratio}}}
  \def\@mathcomposite@#1#2#3#4#5{%
     \vcenter{\m@th\@mathcomposite@option
        \dimen@\f@size\p@\dimen@#1\dimen@\dimen@#5\dimen@
        \divide\dimen@ 18
        \edef\@mathcomposite@skipamount{\the\dimen@}%
        \ialign{\hfil$#4##$\hfil\cr
           #2\crcr
           \noalign{\vskip\@mathcomposite@skipamount}%
           #3\crcr}}}
  \makeatother
\newcommand{\agt}{\mathcomposite{\mathrel}{>}{\sim}}
\newcommand{\alt}{\mathcomposite{\mathrel}{<}{\sim}}

\newcommand{\Eqn}[1]{Eq.~(\ref{#1})}  
\newcommand{\Fig}[1]{FIG.~{\ref{#1}}}
\newcommand{\beq}{\begin{equation}}
\newcommand{\eeq}{\end{equation}}
\newcommand{\ba}{\begin{array}}
\newcommand{\ea}{\end{array}}
\newcommand{\bea}{\begin{eqnarray}}
\newcommand{\eea}{\end{eqnarray}}
\newcommand{\bal}{\begin{align}}  
\newcommand{\eal}{\end{align}}
\newcommand{\bi}{\begin{itemize}}  
\newcommand{\ei}{\end{itemize}}
\newcommand{\ben}{\begin{enumerate}}  
\newcommand{\een}{\end{enumerate}}

\newcommand\hide[1]{}

\newcommand{\ie}{{i.e.}}

\newcommand{\ds}[1]{
  \setbox0=\hbox{\ensuremath{#1}}
  \hbox to\wd0{\hbox to0pt{\hbox to\wd0{\hss/\hss}\hss}\box0}}

\newcommand{\MeV}{\,{\rm MeV}}

\begin{document}

\begin{flushright}
BARI-TH/07-574
\end{flushright}
\vspace*{-5ex}

\title{Possible crossover from BCS superconductivity to Bose-Einstein
condensate in quark matter}


\author{Hiroaki Abuki}{
  address={INFN - Sezione di Bari, Via Amendola 173, I-70126 Bari,
  Italy%
 \\email:~{\rm\texttt{hiroaki.abuki@ba.infn.it}}%
  },
}

\begin{abstract}
The possibility of the crossover from the BCS pairing to the
 Bose-Einstein condensate (BEC) of diquarks with going down
 in density is discussed in the framework of in the Nambu Jona-Lasinio
 (NJL) model.
We find that the quark matter at moderate density may be close to
 the intermediate of the crossover, the precursory regime to the BEC
 phase.
\end{abstract}

\maketitle

\section{Introduction}
It is now established that the ground state of QCD at extremely high
baryon density is in the color-flavor locked superfluid \cite{reviews}
where quarks with all the three flavors participate in the Cooper
pairing.
The appearance of the superfluid CFL is completely due to the BCS
mechanism saying that any arbitrary weak attraction between quarks
leads to the tachyonic Cooperon on the Fermi surface at sufficiently low
temperature.

For the compact star phenomenology, however, one needs to consider
other candidates with less symmetric pairing because the strange
quark mass is not so much smaller compared to quark chemical potential
which is at most of order $500\MeV$ even at the center of stars.
The finite value of strange quark mass causes a stress on the CFL phase. 
This kinematic effect is known to bring about a rich variety of phases
at moderate densities; see \cite{Abuki:2004zk,Ruester:2005jc} for the
NJL model studies of QCD phase diagram.

In addition to such kinematic effect, there is the other key ingredient
which plays an important role at low densities;
that is the strong coupling nature due to the asymptotic freedom of QCD.
Going down in density, quark-gluon interaction becomes large, and this
may lead a modification in superconducting phases.
In fact, the ladder QCD calculation of the coherence length
indicates that the Cooper pair size decreases significantly toward low
density and it could be of order of inter-quark spacing at $\mu=500\MeV$
\cite{Matsuzaki:1999ww,Abuki:2001be}.
This strongly suggests the crossover from the BCS-type weak coupling
superconductivity to the Bose-Einstein Condensate (BEC) of tightly bound
quark pairs
\cite{Nishida:2005ds,Abuki:2006dv,nakano,He:2007kd,He:2007yj}.
Another possible interesting phenomenon is the formation of pseudogap
above the critical temperature \cite{Kitazawa:2001ft}.
These two scenarios indicate the existence of non-trivial (non-Wigner)
phase above the critical temperature at strong coupling.

Major studies done in past mainly concentrate on the spectral
analysis of collective modes.
If the fluctuation is so large, however, there must be its
feedback to the thermodynamics, the equation of state, for example.
Such strong modification of thermodynamics due to fluctuations may
bring about remarkable astrophysical consequences.

In this short article, after briefly summarizing the application of the
Nozi{\`e}res Schmitt-Rink (NSR) theory \cite{nozieres} to the NJL model
following our detailed analysis \cite{Nishida:2005ds,Abuki:2006dv},
we discuss the relativistic BCS-BEC crossover paying a particular
attention to the significance of the fluctuation feedback to the quark
matter thermodynamics.

\section{Application of the Nozi{\`e}res
 Schmitt-Rink theory to the Nambu Jona-Lasinio model}
We here consider a general relativistic four-fermion model with
a point attraction \cite{Nishida:2005ds},
\beq
 \mathcal{L}[\psi,\bar\psi]
 =\bar\psi\left(i\ds\partial-m+\gamma_0\mu\right)\psi%
   +G\psi^\dagger i\gamma_5C
 \psi^*\psi^\mathrm{T}i\gamma_5C\psi,
\label{eq:lagrangian}
\eeq
where $\psi$, $m$ denote the Dirac spinor field and its mass, 
$\mu$ 
is the chemical potential to adjust the asymmetry between
particle and antiparticle, and $G$ parameterizes the strength of
attraction.
The extension to the NJL model with isospin doublet, three colors is
straightforward and the results will be given later.
After introducing Hubbard-Stratonovich fields $\Delta(\tau,{\mathbf x})$
for $i\psi^\mathrm{T}\gamma_5C\psi$, the fermion can be integrated out:
\beq
 Z=Z_0\int\mathcal D\Delta\mathcal D\Delta^{\!*}\,
 \exp\left(-S_\mathrm{eff}[\Delta,\Delta^{\!*}]\right).
 \label{eq:partitionfunction}
\eeq
$Z_0=e^{-\beta\Omega_0(\mu,T)}$
is the free fermion part of the partition function, while
$S_\mathrm{eff}[\Delta,\Delta^{\!*}]$
is the effective action for the collective bosonic fields.
According to \cite{nozieres}, we include the effect of fluctuation up to
the second order in $\Delta$. We have
$ S_\mathrm{eff}[\Delta,\Delta^{\!*}]%
 =T\sum_{n}\int\frac{d{\mathbf p}}{(2\pi)^3}%
 \left[\frac1{G}-\chi(i\omega_n,{\mathbf p})\right]%
 \bigl|\Delta(i\omega_n,{\mathbf p})\bigr|^2 $ 
where $\omega_n$ is fermionic Matsubara frequency.
Then the field $\Delta$ can be integrated out and the
pressure of the system leads to
\beq
  p(\mu,T)=p_0(\mu,T)+p_{\rm fluc}(\mu,T),
\eeq
where $p_0(\mu,T)=2T\sum_{\sigma=\pm}\int\frac{d\mathbf p}{(2\pi)^3}%
\ln\big(1+e^{-{|\sqrt{\mathstrut p^2+m^2}-\sigma\mu|}/{T}}\big)$ 
is the free fermion contribution, and $p_{\rm fluc}$ corresponds
to the fluctuation contribution defined
by
\beq
 p_{\rm fluc}(\mu,T)=T\sum_{N}\int\frac{d\mathbf K}{(2\pi)^3}%
 \ln\left(1-G\chi_{\mu,T}(i\Omega_N,{\mathbf K})\right).
\eeq
Here, $\chi_{\mu,T}(i\Omega_N,{\mathbf K})={\rm F.T.}\langle T_{\tau}%
\left[\psi^Ti\gamma_5 C\psi(\tau,{\mathbf x})\right]%
\left[\psi^Ti\gamma_5 C\psi(0,{\mathbf 0})\right]^\dagger\rangle$
is the Cooperon at one loop, and $\Omega_N$ is bosonic frequency.
(See \cite{Nishida:2005ds,Abuki:2006dv,Kitazawa:2001ft} for the explicit
expression.)
Note that, since we are approaching $T_c$ from above, the overall factor
$1/2$
is dropped because the phase and amplitude fluctuation contribute
equally to the partition function.

When the temperature (chemical potential) is decreased (increased) from the
normal phase, the Cooperon becomes tachyonic at some critical point.
The thermodynamic stability requires $1-G\chi_{\mu,T}(0,{\mathbf K})>0$.
Because the function takes minimum at $K=0$ for the system without
density imbalance, the condition of criticality is simply
\beq
  1-G\chi_{\mu_c,T_c}(0,{\mathbf 0})=0,
\label{thouless}
\eeq
which is nothing but the Thouless criterion. 
This condition generates a one-dimensional line in the $(\mu,T)$-plane
which we call the Thouless line.
To see how large the fluctuation effect on the thermodynamics is, it is
better to move on to the canonical ensemble.
This corresponds to determine the value of $\mu_c$ by means of
\beq
 \frac{\partial P_0}{\partial\mu}(\mu_c,T_c)+\frac{\partial P_{\rm
 fluc}}{\partial\mu}(\mu_c,T_c)=\frac{k_F^3}{3\pi^2},
\label{number0}
\eeq
where $k_F$ parameterizes the total density of the system.
The first term is the free fermion contribution while the second term is
the fluctuation contribution. 
We will see later that, the second term gives a significant contribution
even in the BCS side.

From above two basic equations, we numerically obtain $T_c$ and $\mu_c$ 
as a function of density and coupling, \ie, $T_c(G,k_F,\Lambda)$ and
$\mu_c(G,k_F,\Lambda)$
where $\Lambda$ is an appropriate momentum cutoff.
Before going into numerical computations, let us briefly
summarize the effect of including $N_c=3$ colors and $N_f=2$ flavors.
If we assume the attraction in the isoscalar and color anti-triplet
channel, the Thouless criterion \Eqn{thouless} is not affected.
On the other hand, the number condition \Eqn{number0} is modified as
follows.
\beq
  N_fN_c\frac{\partial P_0}{\partial\mu}(\mu_c,T_c)%
  +\frac{N_c(N_c-1)}{2}\frac{\partial P_{\rm
  fluc}}{\partial\mu}(\mu_c,T_c)=N_fN_c\frac{k_F^3}{3\pi^2},
\label{number}
\eeq
where $k_F$ is redefined so that each fermion species has
a density $\frac{k_F^3}{3\pi^2}$ when the interaction is turned off.
We see that the fluctuation contribution is multiplied by a kinematic
factor $d_B\equiv\frac{N_c(N_c-1)}{2}$, representing the fact that the
system has $d_B$ collective modes belonging to the
antisymmetric representation of SU${}_{\rm c}(3)$.
From the parametric dependence, 
we see that the fluctuation dominates the thermodynamics in the
$N_c\to\infty$
limit.

In numerical calculations, we set $\frac{mc}{\hbar\Lambda}=0.2$. 
In addition, we mainly study the crossover detail for a {\em relativity
parameter}, $\frac{\hbar k_F}{mc}=0.2$
\footnote{Note that even in the known strong coupling system,
the nuclear matter consisting of neutron and proton, this
parameter is of order $\frac{\hbar k_F}{mc}={{\mathcal
O}(10^{-3})}$ at the Mott transition point \cite{Lombardo:2001ek}.},
leaving to see its density/mass dependence later.
Further we use the modified coupling $g=\frac{G_{Rc}}{G_R}$ instead
of bare coupling $G$ \footnote{The regularized coupling $G_R$ 
is introduced by $-\frac{1}{G_R}=\frac{1}{G}-\chi_{0,0}(2m,0)$,
and $G_{Rc}$ stands for the critical regularized coupling for zero mass
boson at vacuum, \ie,
$-\frac{1}{G_{Rc}}=\chi_{0,0}(0,0)-\chi_{0,0}(2m,0)$. $G_R$
is related to the scattering length $a_s$ by $G_R=\frac{4\pi a_s}{m}$.
Then we see that
the zero-binding bound state forms at $g=0$ (the unitary limit) and it
becomes massless at $g=1$ at vacuum.
See \cite{Abuki:2006dv} for the detail.}.

\begin{figure}
  \centering
  \includegraphics[width=0.8\textwidth]{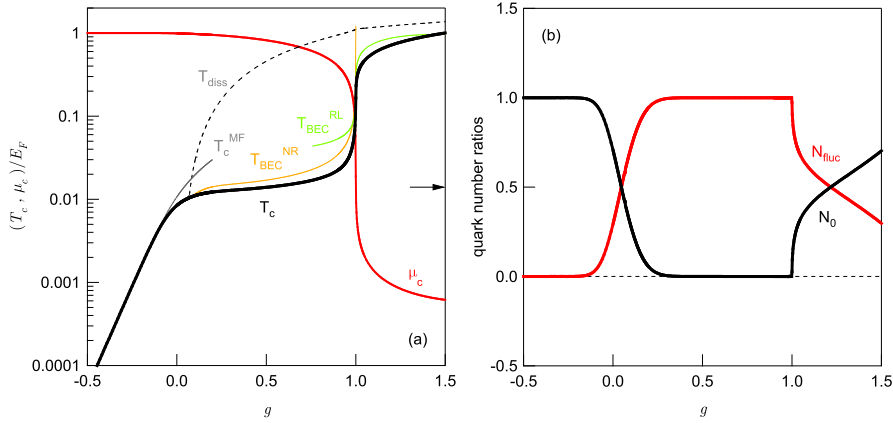}
  \caption{(a)~$T_c$ and $\mu_c$ as a function of the modified coupling
   $g$. 
   These are shown in the unit of the Fermi energy,
   $E_F\equiv\sqrt{m^2+k_F^2}$.
   For other lines, see text.
  (b)~The quark number contents;
  the line denoted by $N_0$ ($N_{\rm fluc}$) corresponds
  to the ratio of the mean field (fluctuation) contribution to
  the total density.
  \vspace*{-2ex}
  }
  \label{fig0}
\end{figure}

In Fig.~\ref{fig0}(a), we show $T_c$, $\mu_c$ as a function of $g$.
When $g\alt -0.2$, $T_c$ is well approximated by
the mean field result indicated by $T_{\rm MF}$.
In the mean field approximation (without $P_{\rm fluc}$), 
$\mu_{\rm MF}\cong E_F$
by neglecting a tiny correction of order $T^2/E_F$.
As a consequence, $T_{\rm MF}$ is determined almost by the
Thouless criterion.
As $g$ is increased and the unitary point $g=0$ is approached, the mean
field result starts to deviate from real $T_c$.
This means that fluctuation contribution in the number equation
\Eqn{number} grows gradually and it cannot be ignored anymore.
In fact, the quark density coming from fluctuation grows as seen in
fig. (b).
When the coupling exceeds $g=0.07$, $\mu_c$ gets lower than fermion
mass $m$ and the in-medium bound state appears accordingly. (See
\cite{Nishida:2005ds,Abuki:2006dv} for the detailed spectral analysis.)
Then the system goes into the BEC phase where $T_c$ is nearly saturated
to a constant and $T_{\rm MF}$ completely fails.
This saturating behaviour suggests that $T_c$ is determined by
the number equation \Eqn{number} because it does not explicitly depend
on $G$.
In fact, the in-medium boson mass is shown to be 
twice of $\mu_c$ ($M_B=2\mu_c$) at $T=T_c$, and 
provided $M_B\gg T_c$ (nonrelativistic),
\Eqn{number} can be approximated by
\beq
  N_cN_f\frac{k_F^3}{3\pi^2}\sim\frac{N_c(N_c-1)}{2}%
  \int\frac{d\mathbf K}{(2\pi)^3}%
  \frac{2}{e^{{K^2}/{2M_B T_c}}-1}\equiv 2d_B%
  \zeta\left({\textstyle\frac{3}{2}}\right)%
  \Big(\frac{M_B T_c}{2\pi}\Big)^{\frac{3}{2}}.
\eeq
The factor 2 in the integrand comes from the fact one diquark
consists of two quarks.
This gives an approximation
$T_c\sim T_{\rm BEC}^{\rm NR}=%
\frac{2}{\pi^{1/3}}\big(\frac{N_c-1}{N_f}%
\zeta(\frac{3}{2})\big)^{-\frac{2}{3}}\frac{k_F^2}{M_B}$. 
In fig. (a), this formula with $M_B=2\mu_c$ is tried by the thin line,
which agrees very well with real $T_c$.
The rightarrow indicates the nonrelativistic strong coupling limit of
$T_c$ , 
evaluated by $T_{\rm BEC}^{\rm NR}$ with $M_B=2m$.
In the current framework, $T_c$ does not saturate to this value and
slightly increases due to the binding effect. 
This is a residual relativistic effect that the binding energy of
diquarks can become as large as the order of the constituent fermion
mass.

As $g$ is increased and $g\cong 1$ is approached, $T_{\rm BEC}^{\rm NR}$
starts to fail and the system eventually goes into the new regime, the
relativistic BEC (RBEC) \cite{kapusta}.
This is because the nonrelativistic approximation $2\mu_c\gg T_c$
is no longer valid there due to the large binding effect.
In fact, the diquark mass is smaller than temperature, $2\mu_c\alt T_c$, 
for $g\agt 1$, and therefore anti-diquarks contribute to the
thermodynamics.
By only taking the stable boson and antiboson contributions in
\Eqn{number}, we get an approximation $T_c\sim T_{\rm BEC}^{\rm RL}%
=\frac{1}{\pi}\sqrt{\mathstrut\frac{k_F^3}{M_B}\frac{N_f}{N_c-1}}$
in the same way as \cite{kapusta}. 
This formula with $M_B=2\mu_c$ is tried by thin line in
\Fig{fig0}(a), which fairly agrees with the real $T_c$.
However, in contrast to the RBEC of the elementary boson \cite{kapusta},
our composite boson system has fermionic degrees of freedom 
due to the competition between the internal energy and entropy
\cite{Abuki:2006dv}.
For this reason, the agreement is not so good as that in the BEC
region.
\Fig{fig0}(b) confirms this picture; quark density from the fermion
sector again comes to play a major role in the deep RBEC region.

Remarkably, in the (R)BEC region, there is a non-trivial phase above
$T_c$, 
the preformed boson phase, up to $T_{\rm diss}$. $T_{\rm diss}$, 
shown by the dashed line in \Fig{fig0}(a), characterizes the ionization
of diquarks.
Needless to say, the system in the preformed boson phase differs much
from a pure Fermi gas although the symmetry is restored.

\begin{figure}
  \centering
  \includegraphics[width=0.8\textwidth]{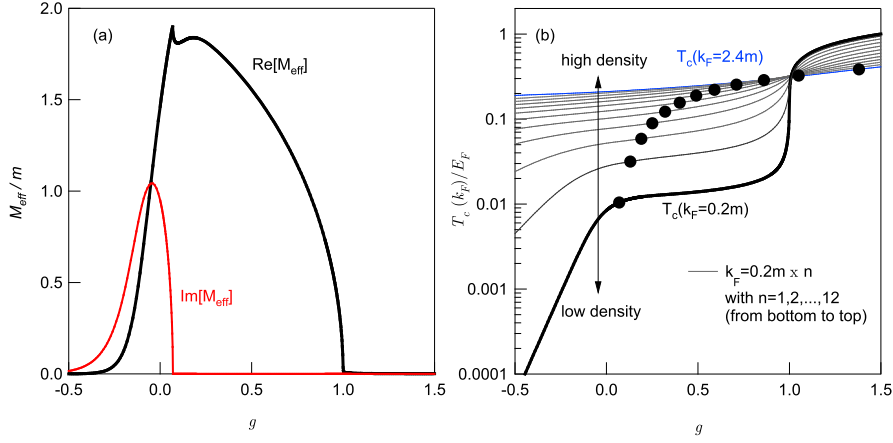}
  \caption{
  (a)~The effective mass $M_{\rm eff}$ for the fluctuation pair field
  as a function of $g$.
  (b)~The change of $T_c$ with respect to the increase of the relativity
  parameter, $\frac{\hbar k_F}{mc}=0.2\times n$ with
  $(n=1,2,\cdots,12)$.
  \vspace*{-2ex}
  }
  \label{fig1}
\end{figure}

Interestingly enough, the definite crossover appears also in the dynamic
equation for the pair excitaion \cite{Abuki:2006dv}.
By expanding $S_{\rm eff}$ up to the quartic order in $\Delta$ and
performing the low energy/long wavelength expansion, we obtain the
dynamic equation near $T_c$
\beq
   -id\frac{\partial}{\partial t}\Delta(t,{\mathbf x})%
   =-\frac{\delta F_{\rm eff}%
   [\Delta,\Delta^*]}{\delta\Delta^*(t,{\mathbf x})}%
   =\left[a_0\frac{T_c-T}{T_c}+\frac{c}{4m}\nabla_{\mathbf x}^2%
   -b_0|\Delta(t,{\mathbf x})|^2\right]\Delta(t,{\mathbf x}),
\eeq
where $d$ (complex) and $\{a_0, c, b_0\}$ (real) are the low energy
coefficients. 
If we define the complex effective mass by $M_{\rm eff}=\frac{2md}{c}$,
its real part coming from the particle-hole asymmetry represents the
propagating piece of the fluctuating pair field \cite{Kitazawa:2007im},
while the imaginary part expresses its diffusive nature.
The real and imaginary parts of $M_{\rm eff}$ as a function of $g$ are
depicted in \Fig{fig1}(a).
In the BCS regime, the pair mode is diffusive, but the magnitude of
damping decreases significantly as the unitary point $g=0$ is
approached.
When the system goes into the BEC phase, the imaginary part vanishes
and fluctuation becomes a pure propagating mode due to a bound state
gap.
However $M_{\rm eff}$ does not saturate to $2m$ in contrast to the
nonrelativistic calculation \cite{haussmann}; it gets smaller towards
the RBEC phase, which is also due to the relativistic
binding effect.

Fig.~\ref{fig1}(b) shows how the crossover characteristics of $T_c$ is
affected by the increase of relativity parameter $\frac{\hbar k_F}{mc}$.
This corresponds to decreasing $m$ or, increasing density $k_F$.
From bottom to top, $\frac{\hbar k_F}{mc}=0.2\times n$ with
$n=1,2,\cdots,12$.
Several notes are in order.
(i) The BCS/BEC crossover point $m=\mu_c$ indicated by the large point
 shifts to higher value of $g$.
 This is due to the Pauli-blocking effect which prevents the formation
 of in-medium bound state at high (small) density (mass).
 At the same time, the crossover characteristics of $T_c$ gets smeared.
(ii) The universal thermodynamics of the unitarity point is absent
 in the relativistic system as noticed in \cite{He:2007kd}. $T_c$ 
 explicitly depends on the additional parameter $\frac{\hbar k_F}{mc}$.
 In fact, the pressure at zero temperature is deduced to take a form
 $p=\frac{2k_F^3}{15\pi^2}\frac{k_F^2}{2m}f(\frac{1}{k_Fa_s},\frac{\hbar
 k_F}{mc},\frac{\hbar\Lambda}{mc})$ 
 with $f$ denoting a dimensionless function;
 we observe only in the $m\to\infty$ limit the nonrelativistic
 universal behaviour at $\frac{1}{k_F a_s}=0$ is recovered.

Let us finally discuss in which regime the actual quark matter does
exist. The modified coupling $g$ is a function of $\Lambda$ and $m$.
It is easy to see $g\to 1$ with $m\to\infty$. Also $g\to-\infty$
when $m\to 0$ with a natural assumption that a bare coupling
$G\Lambda^2$
is less than $\pi^2$, the critical value for the dynamical Majorana mass
generation at vacuum.
As there should be the point $g=0$ in between,
we conclude that for any fixed value for $G<\pi^2$, the system is
BCS-like for $m\to 0$ and it is BEC-like at sufficiently large $m$.
If we fix $G$ to the usually adopted value, $3/4$ of an appropriate
scaler coupling \cite{Abuki:2004zk,Ruester:2005jc}, $g=0$ corresponds to
$m/\Lambda\cong 0.53$. 
This is much larger mass compared to the current quark masses in
agreement with the recently appeared paper \cite{Kitazawa:2007im}.
We conclude that somewhat exotic conditions must be satisfied to have
the diquark BEC in QCD phase diagram; (i) the diquark coupling
is stronger than expected,
and/or (ii) in-medium quark mass is larger than its perturbative estimate.
Interestingly, two lattice studies, one about $qq$ interaction
\cite{Nakamura:2004ur} and the other for in-medium quasiquark mass
\cite{Petreczky:2001yp} are encouraging.
Also it is worth noting that even if the BEC phase cannot be reached,
the fluctuation feedback to quark matter thermodynamics will be
significant.
In fact, it can be neglected only in the weak coupling limit as noted in
\cite{He:2007yj}.
If so it may bring about a remarkable modification of the structure of
possible quark, or hybrid stars.
Exploring possible BCS/BEC crossover in quark matter with more
realistic situations taken into account, as well as looking for its
astrophysical consequences clearly deserves further investigations.

\vspace*{1ex}
\noindent
This work was supported by a Grant-in-Aid for the 21st Century 
   COE ``Center for Diversity and Universality in Physics''.
The numerical calculations were carried out on Altix3700 at YITP in
 Kyoto University.


\begin{thebibliography}{99}

\bibitem{reviews}
  K.~Rajagopal and F.~Wilczek,
  arXiv:hep-ph/0011333;
  M.~G.~Alford,
  Ann.\ Rev.\ Nucl.\ Part.\ Sci.\  {\bf 51}, 131 (2001);
  T.~Sch\"afer,
  arXiv:hep-ph/0304281;
  D.~H.~Rischke,
  Prog.\ Part.\ Nucl.\ Phys.\  {\bf 52}, 197 (2004).

\bibitem{Abuki:2004zk}
  H.~Abuki, M.~Kitazawa and T.~Kunihiro,
  Phys.\ Lett.\  B {\bf 615}, 102 (2005)
  [arXiv:hep-ph/0412382];
  H.~Abuki and T.~Kunihiro,
  Nucl.\ Phys.\  A {\bf 768}, 118 (2006)
  [arXiv:hep-ph/0509172].

\bibitem{Ruester:2005jc}
  S.~B.~Ruester {\it et al.},
  Phys.\ Rev.\  D {\bf 72}, 034004 (2005);
  D.~Blaschke {\it et al.},
  Phys.\ Rev.\  D {\bf 72}, 065020 (2005)

\bibitem{Matsuzaki:1999ww}
  M.~Matsuzaki,
  Phys.\ Rev.\  D {\bf 62}, 017501 (2000)
  [arXiv:hep-ph/9910541].

\bibitem{Abuki:2001be}
  H.~Abuki, T.~Hatsuda and K.~Itakura,
  Phys.\ Rev.\  D {\bf 65}, 074014 (2002)
  [arXiv:hep-ph/0109013];
  K.~Itakura,
  Nucl.\ Phys.\  A {\bf 715}, 859 (2003).

\bibitem{Nishida:2005ds}
  Y.~Nishida and H.~Abuki,
  Phys.\ Rev.\  D {\bf 72}, 096004 (2005)
  [arXiv:hep-ph/0504083].

\bibitem{Abuki:2006dv}
  H.~Abuki,
  Nucl.\ Phys.\  A {\bf 791}, 117 (2007)
  [arXiv:hep-ph/0605081].

\bibitem{nakano}
  K.~Nawa, E.~Nakano and H.~Yabu,
  Phys.\ Rev.\  D {\bf 74}, 034017 (2006);
  J.~Deng, A.~Schmitt and Q.~Wang,
  arXiv:nucl-th/0611097;
  A.~H.~Rezaeian and H.~J.~Pirner,
  Nucl.\ Phys.\  A {\bf 779}, 197 (2006).

\bibitem{He:2007kd}
  L.~He and P.~Zhuang,
  Phys.\ Rev.\  D {\bf 75}, 096003 (2007)
  [arXiv:hep-ph/0703042].

\bibitem{He:2007yj}
  L.~He and P.~Zhuang,
  arXiv:0705.1634 [hep-ph].

\bibitem{Kitazawa:2001ft}
  M.~Kitazawa {\it et al.}, 
  Phys.\ Rev.\  D {\bf 65}, 091504 (2002);
  Phys.\ Rev.\  D {\bf 70}, 056003 (2004);
  Prog.\ Theor.\ Phys.\  {\bf 114}, 117 (2005).


\bibitem{nozieres}
  P.~Nozi{\`e}res and S.~Schmitt-Rink, 
  J.\ Low.\ Temp.\ Phys.\ {\bf 59} 195 (1985).

\bibitem{Lombardo:2001ek}
  U.~Lombardo {\it et al.},
  Phys.\ Rev.\  C {\bf 64}, 064314 (2001).

\bibitem{kapusta}
  H.E.~Haber and H.A.~Weldon, Phys.\ Rev.\ Lett.\ {\bf 46}, 1497
 (1981); J.I. Kapusta, Phys.\ Rev.\ D {\bf 24}, 426 (1981).

\bibitem{haussmann}
  R.~Haussmann, Phys.\ Rev.\ B {\bf 49}, 12975 (1994).

\bibitem{Kitazawa:2007im}
  M.~Kitazawa, D.~H.~Rischke and I.~A.~Shovkovy,
  arXiv:0707.3966 [nucl-th]; arXiv:0709.2235 [hep-ph].

\bibitem{Nakamura:2004ur}
  A.~Nakamura and T.~Saito,
  Prog.\ Theor.\ Phys.\  {\bf 112}, 183 (2004)
  [arXiv:hep-lat/0406038].

\bibitem{Petreczky:2001yp}
  P.~Petreczky {\it et al.},
  Nucl.\ Phys.\ Proc.\ Suppl.\  {\bf 106}, 513 (2002).

\end{thebibliography}
\end{document}